# Valence Parton Density in the Pion from QCD Sum Rules

A.V.Belitsky

*Bogoliubov Laboratory of Theoretical Physics*
*Joint Institute for Nuclear Research*

**Abstract**

We calculate the leading twist valence quark distribution in the pion in the framework of QCD sum rules with nonlocal condensates. Particular attention has been paid to the correct account for the bilocal power corrections.

The determination of parton distributions is, up to now, reserved to experimental studies but as a final goal they are expected to be evaluated from the first principles of the theory. In the lack of complete understanding of the yet unclear confinement mechanism they provide a challenging task for nonperturbative methods presently available. Among the approaches which account for nonperturbative effects the most close to QCD perturbation theory are the QCD sum rules [1]. In the last decade they were applied with moderate success to determine nucleon and photon structure functions in the region of intermediate values of the Bjorken variable [2] and, recently, in the small $\lambda$-region for the light-cone position representation [3]. While the nucleon structure functions are now well defined by the analyses of the precise experimental data and are attacked theoretically, much less is known about the parton distribution of other hadrons, in particular of $\pi$-meson. Being of interest in their own right they provide good testing ground for predictions of the QCD sum rule method.

In order to evaluate the quark distribution in the pion by means of the QCD sum rules method, we consider an appropriate three-point correlation function

$$W_{\mu\nu}(p_1, p_2, q) = i^2 \int d^4\mathrm{x} d^4\mathrm{y} e^{ip_1\mathrm{x}+iq\mathrm{y}} \langle 0|T\left\{j_\mu^5(\mathrm{x}), \mathcal{O}\left(\mathrm{y}+\frac{\lambda}{2}n, \mathrm{y}-\frac{\lambda}{2}n\right), j_\nu^{5\dagger}(0)\right\}|0\rangle. \tag{1}$$

of two axial currents that have non-zero projection onto the pion state being proportional to the pion decay constant $\langle 0|j_\mu^5|\pi(p)\rangle = if_\pi p_\mu$ and the nonlocal string operator $\mathcal{O}$ on the light cone defined by equation

$$\mathcal{O}\left(\frac{\lambda}{2}n, -\frac{\lambda}{2}n\right) = \bar{u}\left(\frac{\lambda}{2}n\right)\gamma_+ \Phi\left[\frac{\lambda}{2}n, -\frac{\lambda}{2}n\right] u\left(-\frac{\lambda}{2}n\right) + (\lambda \to -\lambda) \tag{2}$$

and $\Phi$ is a path ordered exponential in the fundamental representation of the colour group along the straight line $C$ which insures the gauge invariance of the parton distribution. This operator when sandwiched between the pion states defines the twist-2 valence quark distribution in a hadron [4]

$$\langle h(p)|\mathcal{O}\left(\frac{\lambda}{2}n, -\frac{\lambda}{2}n\right)|h(p)\rangle = 4\int_0^1 \cos(\lambda x) u_h(x). \tag{3}$$

hep-ph/9612455  24 Dec 1996

The usual strategy is to use the duality between the hadronic and partonic representation for the correlator under investigation.

On the one hand, we should consider the dispersion relation for the latter and extract the contribution due to the low lying hadron, namely, due to $\pi$-meson, approximating the higher state contribution by perturbative spectral density. We perform the double Borel transformation and put the parameters equal $M_1^2 = M_2^2 = 2M^2$ in order not to introduce the asymmetry between the initial and final pion states and to make contact with two-point sum rules for the pion decay constant.

On the other hand, we consider the OPE for the same quantity. Of course, the QCD sum rules with local condensates are inappropriate here because the usual local power corrections produce $\delta$-type contribution to the distribution function. It is not surprising since some propagators are substituted by constant factors that do not allow the momentum to flow and the whole hadron momentum be carried by a single quark. The probability density of this configuration in the phase space is $\delta(1-x)$. Higher condensates produce even more singular terms. However, this singular contribution can be smeared over the whole region of the momentum fraction from zero to unity by avoiding the Taylor expansion of the generic nonlocal objects which are the starting point of all QCD sum rule calculations and introducing the concept of nonlocal condensate which assumes the finite correlation length for the vacuum fluctuations.

At the two-loop level, to which we restrict our analysis, we need the bilocal quark and gluon condensates, trilocal quark-gluon condensates and four-quark condensates. The latter will be factorized into the product of bilocal scalar quark condensates via the vacuum dominance hypothesis. For explicit calculations, it is convenient to parametrize the bilocal condensates in the form of the well-known $\alpha$-representation for propagators [5]

$$\langle 0|\bar\psi(0)\Phi[0,\mathrm{x}]\psi(\mathrm{x})|0\rangle = \langle\bar\psi\psi\rangle \int_0^\infty d\alpha f_S(\alpha) e^{\alpha \mathrm{x}^2/4},$$

$$\langle 0|\bar\psi(0)\Phi[0,\mathrm{x}]\gamma_\mu\psi(\mathrm{x})|0\rangle = -i\mathrm{x}_\mu \frac{2}{81}\pi\alpha_s\langle\bar\psi\psi\rangle^2 \int_0^\infty d\alpha f_V(\alpha) e^{\alpha \mathrm{x}^2/4}. \qquad (4)$$

We use the following ansatz for the distribution of vacuum quarks in the virtuality $\alpha$ [6]

$$f_S(\alpha) = \frac{\sqrt{\gamma}}{2\Lambda K_1(2\Lambda\sqrt{\gamma})} \exp\left(-\frac{\Lambda^2}{\alpha} - \alpha\gamma\right). \qquad (5)$$

Here $\Lambda^2 = 0.2 GeV^2$, and $\gamma$ is fixed from the lowest nontrivial moment of the distribution function $f_S$ that is related to the value of the average virtuality of the vacuum quarks $\lambda_q^2$.

Conventional calculations of the perturbative diagram with the light quark masses neglected result in

$$\Psi_{pert}(M^2, Q^2, \lambda) = \frac{3}{2\pi^2} M^2 \int_0^1 dx \cos(\lambda x) x\bar{x} \exp\left(-\frac{\bar{x}}{x}\frac{Q^2}{4M^2}\right). \qquad (6)$$

Note that we have kept the $t$-channel momentum transferred to be nonzero. If we expand the cosine in the Taylor series and integrate over $x$, we find out that each moment possesses logarithmic non-analyticities of the type $(Q^2)^n \ln Q^2$. These terms come from the small-$x$ region, where the spectator quark carries almost the whole momentum of the pion, so that the struck quark becomes wee and can propagate over large distances in the $t$-channel. Therefore, we have to perform additional factorization for separation of small and large distances in the corresponding invariant amplitude; this will lead to the appearance of additional terms in the OPE for the three-point correlation function which correct the small-$x$ dependence of the parton density.

The simplest nonperturbative correction comes from the vector condensate

$$\Psi_V(M^2, \lambda) = \frac{8}{81}\pi\alpha_s\langle\bar{u}u\rangle^2 \int_0^1 dx \cos(\lambda x) x f_V(\bar{x}M^2). \tag{7}$$

The dominant contribution is due to the four-quark condensate. Performing straightforward calculations we obtain

$$\Psi_S(M^2, \lambda) = \frac{32}{9}\pi\alpha_s\langle\bar{u}u\rangle^2 \int_0^1 dx \cos(\lambda x)$$
$$\times \int_0^1 dy \int_0^1 d\xi \int_{\frac{1}{2}}^1 d\zeta f_S\left(\frac{\bar{x}}{\xi}M^2\right) f_S\left(\frac{y}{\zeta}M^2\right) \theta\left(\frac{\xi - \zeta}{x - y}\right) \frac{x\bar{y}}{|\bar{x}y\bar{\xi}\zeta - x\bar{y}\xi\bar{\zeta}|}. \tag{8}$$

The gluon as well as trilocal quark-gluon condensate contributions are numerically much less important than the power correction we accounted for; therefore, we neglect them in what follows.

It is well known that there exists a parton sum rule that implies that the pion contains one $u$-quark. Summing the calculated contributions and taking the formal limit $Q^2 \to 0$ in the perturbative term we can convince ourselves, comparing the result with the sum rule for the pion decay constant [5], that the normalization condition is broken. The reason for this has already been mentioned earlier and we elaborate this point below.

As we have seen, in the limit $Q^2 \to 0$ the perturbative term though finite contains the logarithmic non-analyticities at this point. This is a typical example of the mass singularities in the QCD sum rules framework [7, 8, 9]. In order to get rid of this perturbative behaviour and replace it by a physical one, it is necessary to modify the original OPE. For the form factor type problem a two-fold structure of the modified OPE has been realized in refs. [10, 11] being of the following schematic form:

$$W(p_1^2, p_2^2, q^2) = \sum_d C^{(d)}(p_1^2, p_2^2, q^2)\langle\mathcal{O}_d\rangle + \sum_i \int d^4\mathrm{x} e^{ip_1\mathrm{x}} \mathcal{C}^{(i)}(\mathrm{x}) \mathcal{W}^i(q, \mathrm{x}, \lambda). \tag{9}$$

An additional second term determines the contribution due to the long-distance propagation of quarks in the $t$-channel. Here $\mathcal{W}^i$ are the two-point correlators of the operator in question and some nonlocal string operator of a definite twist [9] which arises from the OPE of $T$-product of pion interpolating fields. The coefficients $C^{(d)}(p_1^2, p_2^2, q^2)$ in eq. (9) are free from non-analyticities or singularities in $Q^2$ because they are defined as the difference between the original diagram and its factorized expression which is the perturbative analogue of the corresponding bilocal correlator.

The simplest bilocal power correction is given by the convolution of the quark propagator $\mathcal{C}^{(1)}(\mathrm{x}) = 2S_+(\mathrm{x})$ and

$$\mathcal{W}^V_{++}(q, \mathrm{x}, \lambda n) = i \int d^4\mathrm{y}\, e^{iq\mathrm{y}} \langle 0|T\left\{\bar{u}(0)\gamma_+\Phi[0,\mathrm{x}]u(\mathrm{x}), \mathcal{O}\left(\mathrm{y}+\frac{\lambda}{2}n, \mathrm{y}-\frac{\lambda}{2}n\right)\right\}|0\rangle. \quad (10)$$

We extract the contact term due to the vector condensate from this correlator and saturate the remaining part by the contributions of the mesons of increasing spin; these are $\rho^0$, $g$ states and so on. The net result for the difference between the "exact" bilocal and its perturbative part reads

$$\Psi^{(1)}_{BL}(M^2, \lambda) = \frac{8}{81}\pi\alpha_s\langle\bar{u}u\rangle^2 \int_0^1 dx\, \cos(\lambda x)\bar{x}f_V(xM^2) + Q^2 e^{\frac{Q^2}{4M^2}} \sum_{J=1,3,\ldots}^\infty (i\lambda)^{J-1} \quad (11)$$

$$\times\left\{\frac{3}{8\pi^2}\frac{1}{\Gamma(J)\Gamma(J+2)}\int_0^{z_J^0} dz\, \frac{z^J}{z+\frac{Q^2}{4M^2}} + (-2)^J \left(\frac{m_M^2}{M^2}\right)^J \frac{f_J^{(1)}f_J^{(2)}}{m_M^2+Q^2}\int_0^1 d\beta\, \varphi_J^{(2)}\left(\frac{1+\beta}{2}\right)\right\}$$

where $z_J^0 = \sigma_J^0/4M^2$, $\sigma_J^0$ is the continuum threshold and $m_M$ is the mass of the lowest meson state in the channel of given spin $J$. Here $\varphi_J^{(2)}$ are the wave functions describing the light-cone momentum fraction distribution of quarks inside mesons that parametrize (to the leading twist accuracy) the matrix elements

$$\langle 0|\bar{\psi}(0)\Phi[0,\mathrm{x}]\gamma_\mu\psi(\mathrm{x})|M^J(q,\eta)\rangle = \epsilon^{(\eta)}_{\mu\mu_1\mu_2\ldots\mu_{J-1}} \mathrm{x}_{\mu_1}\mathrm{x}_{\mu_2}\ldots\mathrm{x}_{\mu_{J-1}}(m_M)^J f_J^{(1)}\phi_J^{(1)}(xq)$$
$$- iq_\mu\epsilon^{(\eta)}_{\mu_1\mu_2\ldots\mu_J} \mathrm{x}_{\mu_1}\mathrm{x}_{\mu_2}\ldots\mathrm{x}_{\mu_J}(m_M)^J f_J^{(2)}\phi_J^{(2)}(xq), \quad (12)$$

where $J$ is a spin of the meson, $\eta$ its polarization and $\epsilon^{(\eta)}_{\mu\mu_1\mu_2\ldots\mu_{J-1}}$ is a polarization tensor.

The former term in eq. (12) is a contact-type contribution due to the vector condensate. The first one in the curly brackets is the difference between the perturbative analogue of the bilocal correlator and the continuum contribution into the "exact" one. This part cancels the logarithmic non-analyticities in the perturbative diagram (eq. (6)) corresponding to the leading twist-2 operator in the OPE of pion currents. The tower of the next-to-leading non-analyticities can be subtracted in a similar way by accounting for the twist-4 operator. The last term displays the physical contribution to the correlation function that possesses the correct behaviour in the "momentum transferred $Q^2$".

The sum of eqs. (6) and (12) is an analytical function in $Q^2$ as all singularities are replaced by the combination $Q^2 + \sigma_J^0$ which is safe in the limit $Q^2 \to 0$. Due to the presence of the non-analyticities in each moment of the distribution function, we need an infinite number of parameters to be found from additional sum rules. Obviously, this is an impossible task. Safely, for the problem at hand, this part vanishes in the forward limit and the sum rule is dominated by the contact terms.

The dominant contribution comes from the bilocal correlator convoluted with a three-propagator coefficient function. All nonperturbative information is accounted for in the correlator

$$\mathcal{W}^S_+(q, \mathrm{x}, \lambda n) = i \int d^4 \mathrm{y}\, e^{iq\mathrm{y}} \langle 0 | T \left\{ \bar{u}(0) \Phi[0, \mathrm{x}] u(\mathrm{x}),\, \mathcal{O}\left(\mathrm{y} + \frac{\lambda}{2} n, \mathrm{y} - \frac{\lambda}{2} n\right) \right\} | 0 \rangle. \quad (13)$$

Extracting the contact term and performing the integration by using the method outlined at the beginning of the paper, we obtain the following contribution to the structure function:

$$\Psi^{(2)}_{BL}(M^2, \lambda) = \frac{32}{9} \pi \alpha_s \langle \bar{u} u \rangle^2 \int_0^1 dx \cos(\lambda x)$$
$$\times \int_0^1 dy \int_0^1 d\xi \int_0^{\frac{1}{2}} d\zeta f_S\left(\frac{\bar{y}}{\xi} M^2\right) f_S\left(\frac{x}{\zeta} M^2\right) \theta\left(\frac{\zeta - \xi}{x - y}\right) \frac{y \bar{x}}{|\bar{x} y \xi \zeta - x \bar{y} \bar{\xi} \zeta|}. \quad (14)$$

Now, having accounted for additional terms in OPE, we can easily check that the normalization condition for the quark distribution in the pion is restored.

For zero $Q^2$ the perturbative spectral density is concentrated on the line $s_1 = s_2$, so that there is no transition between the states with different masses. We collect all contributions and make the continuum subtraction that results in the substitution $M^2 \to M^2(1 - \exp(-s_0/M^2))$ in the perturbative term. We have found good stability of the distribution function with respect to the variation of the Borel parameter in the region $0.5 \leq M^2 \leq 0.8$ for the standard value of the continuum threshold $s_0 = 0.7 GeV^2$. The normalization point of the OPE is $\mu^2 \sim 0.5 GeV^2$; therefore, the function obtained can be regarded as an "input" quark distribution at this low energy scale. In Fig. 1, we present the curves for the valence quark distribution in the pion for $M^2 = 0.6 GeV^2$: the solid and long-dashed lines correspond to the values of the average virtuality of vacuum quarks $\lambda_q^2 = 0.6 GeV^2$ ($\gamma^{-1} = 0.154 GeV^2$) and $\lambda_q^2 = 0.4 GeV^2$ ($\gamma^{-1} = 0.087 GeV^2$), respectively. In the large-$x$ region, the corrections due to the quark condensate do not exceed 30% of the perturbative term. However, in the small-$x$ region at $x = 0.2$ the ratio of the contact term to the main one comprises 50% for $\lambda_q^2 = 0.6 GeV^2$ and 70% for $\lambda_q^2 = 0.4 GeV^2$. Below this point the nonperturbative contribution increases and reaches 100% at $x = 0.13$ for $\lambda_q^2 = 0.4 GeV^2$ (for $\lambda_q^2 = 0.6 GeV^2$ it still amounts 50%). So, for $x$ as small as 0.2 we could not trust $x$-dependence of our result. Of course, there is no possibility to reproduce the correct $x \to 0$ behaviour of the parton density in the present approach as it is determined by the exchanges of the Regge trajectories.

Since our result is valid only in the limited region of Bjorken variable, we could not evolve it to the experimentally accessible energies. In Fig. 1, we compare our calculation with the distribution obtained in the NJL model [12] at the same normalization point and find reasonable agreement between two approaches in a wide region of the momentum fraction. In Fig. 2, the latter evolved up to $Q^2 = 20 GeV^2$ (short-dashed curve) is compared with the presently available fits of experimental data. It shows good agreement with the result of the analysis of Sutton, Martin, Roberts and Stirling (solid curve) [13], which is consistent with all present Drell-Yan and prompt photon $\pi N$ data. We also present the (long-dashed) curve due to Glück, Reya and Vogt [14]; however, their result does not agree with E615 experiment which requires the valence distribution to be larger by 20%. If the GRV curve is renormalized within a factor of $1.2 - 1.3$ in the central region, there will be no disagreement between the different analyses.

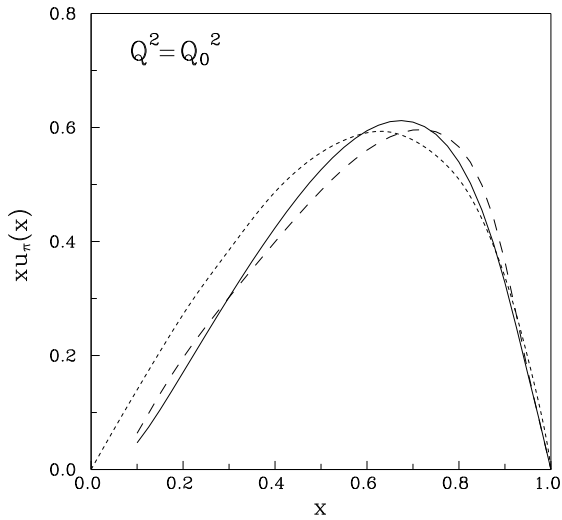
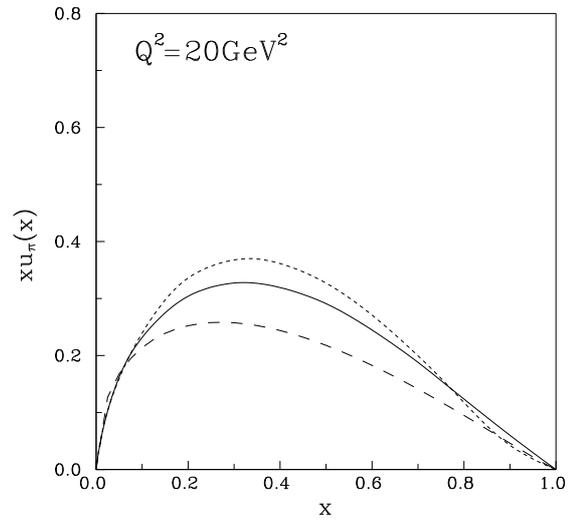

Fig. 1: Quark distribution in the pion at the low energy scale $\mu^2 \sim 0.5 GeV^2$ calculated from the QCD sum rule for different values of the average virtuality of vacuum quarks: solid and long-dashed curves correspond to $\lambda_q^2 = 0.6 GeV^2$ and $\lambda_q^2 = 0.4 GeV^2$, respectively. Short-dashed curve is the $u$-quark density found in the NJL model [12].

Fig. 2: The experimental fits of the valence $u$-quark distribution in the pion at $Q^2 = 20 GeV^2$: SMRS result [13] is depicted by solid curve, GRV analysis [14] is shown by long-dashed one. Short-dashed curve is the $u$-quark density calculated in the NJL model [12] evolved up to $Q^2 = 20 GeV^2$.

In conclusion, we have calculated the pionic parton density at low momentum scale in QCD sum rules with nonlocal condensates. It is shown that the parton sum rule is fulfilled only after the bilocal power corrections are accounted for. We have found good agreement with the $u$-quark distribution function computed in the NJL model which when evolved up to the experimental scales is well comparable with data.

We would like to thank S.V. Mikhailov and R. Ruskov for useful discussions at an early stage of the work, A. Tkabladze for help in numerical calculations and Prof. W.J. Stirling for providing the Fortran package for the evolution of the pion distribution extracted from the experimental data. This work was supported by the Russian Foundation for Fundamental Investigation, Grant $N$ 96-02-17631.

# References

[1] M.A. Shifman, A.I. Vainshtein, V.I. Zakharov, Nucl. Phys. B 147 (1979) 385, 448.

[2] B.L. Ioffe, JETP Lett. 42 (1985) 327, ibid. 43 (1986) 406;
V.M. Belyaev, B.L. Ioffe, Nucl. Phys. B 310 (1988) 548, ibid. B 313 (1989) 647;
Int. J. Mod. Phys. A 6 (1991) 1533;
A.S. Gorsky, B.L. Ioffe, A.Yu. Khodjamirian, A. Oganesian, Z. Phys. C 44 (1990) 523; Sov. Phys. JETP 70 (1990) 25;
A.S. Gorsky, B.L. Ioffe, A.Yu. Khodjamirian, Z. Phys. C 53 (1991) 299;
B.L. Ioffe, A.Yu. Khodjamirian, Phys. Rev. D 51 (1995) 3373;
A.V. Belitsky, J. Phys. G (in press).

[3] V.M. Braun, P. Gornicki, L. Mankiewicz, Phys. Rev. D 51 (1995) 6036.

[4] J.C. Collins, D.E. Soper, Nucl. Phys. B 194 (1982) 445.

[5] S.V. Mikhailov, A.V. Radyushkin, JETP Lett. 43 (1986) 712.

[6] A.P. Bakulev, S.V. Mikhailov, Z. Phys. C 68 (1995) 451.

[7] V.A. Nesterenko, A.V. Radyushkin, JETP Lett. 39 (1984) 707.

[8] V.M. Belyaev, I.I. Kogan, preprint ITEP-29 (1984); Int. J. Mod. Phys. A 8 (1993) 153.

[9] A.V. Radyushkin, R. Ruskov, Phys. Atom. Nucl. 56 (1993) 630, ibid 58 (1995) 1440; Phys. Lett. B 374 (1996) 173.

[10] I.I. Balitsky, Phys. Lett. B 114 (1982) 53;
I.I. Balitsky, A.V. Yung, Phys. Lett. B 129 (1983) 328.

[11] K.G. Chetyrkin, S.G. Gorishny, A.B. Krasulin, S.A. Larin, V.A. Matveev, Preprint IYaI-P-0337 (1984).

[12] T. Shigetani, K. Suzuki, H. Toki, Phys. Lett. B 308 (1993) 383.

[13] P.J. Sutton, A.D. Martin, R.G. Roberts, W.J. Stirling, Phys. Rev. D 45 (1992) 2349.

[14] M. Glück, E. Reya, A. Vogt, Z. Phys. C 53 (1992) 651.